\newif\ifepl \epltrue
\ifepl
\documentclass[doublecol,english]{epl2}
\else
\documentclass[aps,prl,english,twocolumn]{revtex4}
\fi

\usepackage[T1]{fontenc}
\usepackage[latin9]{inputenc}
\usepackage{amsmath}
\usepackage{color}
\usepackage{graphicx}
\usepackage{amssymb}
\usepackage{esint}
\usepackage{wasysym}

\makeatletter
\@ifundefined{textcolor}{}
{%
 \definecolor{BLACK}{gray}{0}
 \definecolor{WHITE}{gray}{1}
 \definecolor{RED}{rgb}{1,0,0}
 \definecolor{GREEN}{rgb}{0,1,0}
 \definecolor{BLUE}{rgb}{0,0,1}
 \definecolor{CYAN}{cmyk}{1,0,0,0}
 \definecolor{MAGENTA}{cmyk}{0,1,0,0}
 \definecolor{YELLOW}{cmyk}{0,0,1,0}
 }

\@ifundefined{definecolor}
 {\usepackage{color}}{}
\@ifundefined{definecolor}{\@ifundefined{definecolor}
 {\usepackage{color}}{}
}{}
\usepackage{epsf}

\newcommand{\bcpo}{BiCu$_2$PO$_6$}
\newcommand{\ybco}{YBa$_2$Cu$_3$O$_{6+\delta}$}
\newcommand{\ybca}{YBa$_2$Cu$_3$O$_{6.6}$}

\newcommand{\lsca}{La$_{1.85}$Sr$_{0.15}$CuO$_{4}$}

\newcommand{\w}{\omega}

\newcommand{\xib}{\xi_{\rm b}}
\newcommand{\xiQ}{\xi_Q}
\newcommand{\xiSG}{\xi_{\rm SG}}
\newcommand{\Nimp}{N_{\rm imp}}
\newcommand{\TN}{T_{\rm N}}
\newcommand{\Tg}{T_{\rm g}}
\newcommand{\Jmax}{J_{\rm max}}
\newcommand{\Nr}{N_{\rm rl}}
\newcommand{\xiconf}{\xi_{\rm b}^{\rm conf}}

\definecolor{maroon}{rgb}{0.5,0.1,0.1}

\makeatother

\usepackage{babel}

\makeatother

\usepackage{babel}

\ifepl
\newcommand{\onlinecite}{\cite}
\newcommand{\prl}{Phys. Rev. Lett. }
\newcommand{\prb}{Phys. Rev. B }
\else
\begin{document}
\fi

\title{
Defect-induced spin-glass magnetism in incommensurate spin-gap magnets
}

\ifepl

\author{Eric C. Andrade\inst{1} \and Matthias Vojta\inst{1}}
\shortauthor{E. C. Andrade and M. Vojta}

\institute{
\inst{1} Institut f\"ur Theoretische Physik, Technische Universit\"at Dresden, 01062 Dresden, Germany
}

\else

\author{Eric C. Andrade}
\author{Matthias Vojta}
\affiliation{Institut f�r Theoretische Physik, Technische Universit\"at Dresden,
01062 Dresden, Germany}

\fi

\date{\today}

\ifepl
\abstract{
We study magnetic order induced by non-magnetic impurities in quantum
paramagnets with incommensurate host spin correlations. In contrast
to the well-studied commensurate case where the defect-induced magnetism
is spatially disordered but non-frustrated, the present problem combines
strong disorder with frustration and, consequently, leads to spin-glass
order. We discuss the crossover from strong randomness in the dilute limit
to more conventional glass behavior at larger doping, and numerically
characterize the robust short-range order inherent to the spin-glass phase.
We relate our findings to magnetic order in both \bcpo\ and \ybca\ induced
by Zn substitution.
}
\else
\begin{abstract}
We study magnetic order induced by non-magnetic impurities in quantum
paramagnets with incommensurate host spin correlations. In contrast
to the well-studied commensurate case where the defect-induced magnetism
is spatially disordered but non-frustrated, the present problem combines
strong disorder with frustration and, consequently, leads to spin-glass
order. We discuss the crossover from strong randomness in the dilute limit
to more conventional glass behavior at larger doping, and numerically
characterize the robust short-range order inherent to the spin-glass phase.
We relate our findings to magnetic order in both \bcpo\ and \ybca\ induced
by Zn substitution.
\end{abstract}
\fi


\ifepl
\begin{document}
\fi

\maketitle


\ifepl
\section{Introduction}
\fi
Magnetic order induced by non-magnetic impurities 
is a remarkable phenomenon which has been observed in a wide variety
of paramagnetic Mott insulators. It can be rationalized assuming dimer
pairing of spins 1/2 in the host magnet: each non-magnetic impurity
introduces one unpaired spin by breaking up a dimer. These impurity
moments mutually interact by the exchange of gapped bulk excitations, with
the interaction falling off exponentially as function of distance.
If the host system features a bipartite lattice with antiferromagnetic
spin correlations, the interactions between the impurity moments are
non-frustrated, and consequently the ground state can be expected
to be a long-range ordered antiferromagnet. This has been verified
both numerically \cite{imada97,yasuda01,wessel01} and experimentally,
with Zn-doped CuGeO$_{3}$ \cite{hase93,manabe98} and Mg-doped TlCuCl$_{3}$
\cite{oosawa02} being prime examples.

Much less is known about impurity-induced magnetism in hosts with
{\em incommensurate} (IC) correlations, realized e.g. in the Zn-doped
spin ladder \bcpo\ \cite{laflo,tsirlin10,alexander10} and also in the
Zn-doped superconductor \ybca, where recent experiments \cite{ybco}
have demonstrated the appearance of static order at an IC wavevector
close to that of the host correlations. Based on the available experimental
data, it has been suggested that the physics is not very different
from the commensurate case.

In this paper, we instead argue that, although the mechanism of moment
formation by impurities is similar in the commensurate and IC cases,
the low-energy physics of these moments is vastly different. The IC case leads
to frustration of the inter-moment interaction and thus to
spin-glass behavior. The actual ordering temperature where spins freeze
into a glassy state is suppressed as compared to the commensurate
case due to frustration, leading to a broad temperature regime of
slow fluctuations. The glassy regime is characterized by pronounced
short-range order at (or near) the wavevector imprinted by the host
correlations, with a correlation length which vanishes as the impurity
concentration $x$ is tuned towards zero. Dimensionality plays an
important role: Glassiness is most pronounced if the host correlations
are IC in three spatial directions, whereas a host with only one-dimensional (1d)
incommensurability leads to a weakly glassy state with large magnetic correlation
length.

In the body of this paper, we present general arguments and extensive
numerical simulation data, comparing the commensurate and IC cases,
which lead to the above conclusions. Our analysis sheds light onto
a class of states which we believe is quite common to moderately disordered
magnets, namely states with robust static short-range order combined
with some amount of glassiness. We connect our results to the concrete
experiments on doped spin-gap magnets in Refs.~\cite{laflo,ybco}.


\ifepl
\section{Effective model}
\else
\textit{Effective model.}
\fi
We shall assume that a mechanism of dimer
breaking underlies the formation of magnetic moments upon doping non-magnetic
impurities. In its most general form, this mechanism relies on confinement
of spinons in the host paramagnet and as such should apply to all
spin-gap magnets with conventional spin-1 (triplon) excitations \cite{vbs00}.

In the dilute limit of small $x$, there is a separation of energy scales, such that the
impurity-induced moments provide the only degrees of freedom below the spin-gap energy
$\Delta$. Hence, the low-energy physics can be captured by an effective spin-$S$
Heisenberg model involving the impurity moments $\vec{S}_{i}$ at random positions $r_{i}$
only,
\begin{equation}
\mathcal{H}_{{\rm eff}}=-\sum_{ij}J_{ij}\vec{S}_{i}\cdot\vec{S}_{j},
\label{heff}
\end{equation}
where $S$ usually equals the bulk spin size. The interaction $J_{ij}$ is dictated by the
bulk magnetic properties. In a static linear-response approximation
$J_{ij}=J_{0}^{2}\chi(\vec{r}_{ij})$ where $J_{0}$ is of order of the bulk exchange
coupling, and $\chi(\vec{r})$ is Fourier transform of the static bulk susceptibility
$\chi(\vec{q})$. Hence, $|J_{ij}|\propto\exp(-r_{ij}/\xib)$ at long distances, where
$\xib\propto c/\Delta$ is the bulk correlation length and $c$ a mode velocity
\cite{sigrist96,lin_foot,wkconf_foot,doretto}.
The separation of energy scales requires $|J_{ij}|\ll \Delta$ which implies $\ell>\xib$
where the average impurity distance $\ell=x^{-1/d}$ with $d=3$.
In this limit, the distribution of energy scales in the model \eqref{heff} becomes broad:
the maximum coupling for each spin, $J_i^{\rm max}={\rm max}_j |J_{ij}|$, displays
a strongly non-Gaussian distribution with large weight at small energies $J_i^{\rm max} \ll J_0$ \cite{1ddiv}.


\ifepl
\section{Low-temperature order}
\else
\textit{Low-temperature order.}
\fi
If the underlying lattice is bipartite and $\chi(\vec{q})$ is peaked at the
antiferromagnetic wavevector $Q=(\pi,\pi,\ldots)$, then $\chi(\vec{r})$ changes sign from
site to site. All interactions can be satisfied classically by antiferromagnetic
order of the $\vec{S}_{i}$, which has been shown to survive even in for 
$S=1/2$ \cite{rsrg1,rsrg2,laflo,wessel01,yasuda01}.

In contrast, if $\chi(\vec{q})$ is peaked at an IC wavevector, the sign changes of
$\chi(\vec{r})$ are irregular, and the model \eqref{heff} is generically frustrated \cite{ic_note}.
Thus, spin-glass behavior is expected, which we will characterize below.

It is worth noting that Eq.~\eqref{heff} is different from the Edwards-Anderson (E-A)
spin-glass model \cite{fischer}:
The latter describes spins on a regular lattice with random couplings, whereas
Eq.~\eqref{heff} features spins at random positions with deterministic couplings
\cite{critsg}. As a result, the $J_{ij}$ in \eqref{heff} display {\em correlated} disorder, in
contrast to uncorrelated disorder usually assumed in the E-A model.
Also, the broad distribution of couplings present in Eq.~\eqref{heff}
is not a property of the standard E-A model.


\ifepl
\section{Monte-Carlo simulations}
\else
\textit{Monte-Carlo simulations.}
\fi
Being interested in qualitative
properties at low temperatures $T$ including the ordered phase, we consider
three-dimensional (3d) systems (with possibly anisotropic couplings)
in the classical limit first, and turn to quantum effects later.
We randomly distribute $\Nimp$ unit vectors $\vec{S}_{i}$ on a
cubic lattice of linear size $L$ with periodic boundary conditions,
yielding an impurity concentration $x=\Nimp/L^{3}$. The effective
interaction $J_{ij}$ is generated from Fourier transforming $\chi(\vec{q})$
which is approximated by the inverse of the gapped bulk triplon dispersion
$\varepsilon\left(\vec{q}\right)$ \cite{sigrist96}:
\begin{equation}
J_{ij}=J_{0}^{2} \frac{1}{L^3} \sum_{\vec{q}}
\frac{e^{i\vec{q}\cdot\vec{r}_{ij}}}{\varepsilon\left(\vec{q}\right)}.
\label{jij}
\end{equation}

We perform classical Monte-Carlo (MC) simulations of the effective
model \eqref{heff}, with single-site updates using a combination
of the heat-bath and microcanonical (or over-relaxation) methods \cite{berg_monte_carlo}.
We consider typically $10^{5}$ MC steps per spin for the measurements,
after discarding $10^{5}$ MC steps for equilibration.
For IC bulk correlations, we expect a (hardly-relaxing) spin-glass
behavior, and to efficiently sample all spin configurations 
we employ the parallel-tempering algorithm \cite{partemp_jpsj}.
Disorder averages are taken over $\Nr$ impurity configurations,
with $\Nr=400$ for $L=16$ to $\Nr=50$ for $L=40$.

From the MC data, we calculate thermodynamic observables, e.g., the
specific heat $C$, as well as the neutron-scattering static structure
factor
\begin{equation}
S(\vec{q})=\frac{1}{\Nimp}\big[\sum_{i,j}\left\langle \vec{S}_{i}\cdot\vec{S}_{j}\right\rangle e^{-i\vec{q}\cdot\vec{r}_{ij}}\big]_{av},
\label{sfactor}
\end{equation}
where $\left\langle \cdots\right\rangle $ denotes MC average and
$\left[\cdots\right]_{av}$ average over disorder, and the sum runs
over the $\Nimp$ impurity sites. In a long-range ordered phase
$S\left(\vec{q}\right)/\Nimp\rightarrow M^2 \delta_{\vec{q},\vec{Q}}$,
where $\delta$ is the Kronecker delta and $\vec{Q}$ is the ordering
wavevector. Note that $M\to1$ as $T\to0$, because the classical
ground state is fluctuationless. The behavior of $S(\vec{q})$ near
$\vec{Q}$ allows to extract the correlation length $\xiQ$ characterizing
the {\em static} impurity order ($\xiQ$ is distinct from $\xib$
\cite{xi_note}). To study signatures of spin-glass freezing, we
also consider the spin-glass order parameter
\begin{equation}
q^{\alpha,\beta}\left(\vec{q}\right)=\frac{1}{\Nimp}\sum_{i}S_{i}^{\alpha\left(1\right)}S_{i}^{\beta\left(2\right)}e^{i\vec{q}\cdot\vec{r}_{i}},
\label{q_SG}
\vspace*{-5pt}
\end{equation}
where $\alpha$ and $\beta$ are spin components, and $^{(1)}$ and $^{(2)}$
denote two identical copies of the system (``replicas'') containing the same impurity
configuration. The spin-glass susceptibility is then defined as
\begin{equation}
\chi_{\rm SG}\left(\vec{q}\right)=\Nimp\sum_{\alpha,\beta}\big[\left\langle \left|q^{\alpha,\beta}\left(\vec{q}\right)\right|^{2}\right\rangle \big]_{av}.
\label{chi_sg}
\end{equation}
From $\chi_{\rm SG}(\vec{q})$ near $\vec{q}=0$ we extract a spin-glass
correlation length $\xiSG$. As $q^{\alpha,\beta}$ acquires long-range
order in a spin-glass state, a divergence of $\xiSG$ signals the
onset of spin-glass order, i.e., freezing.

In our finite-size simulations, the ordering (or freezing) temperature
$\Tg$ is most efficiently extracted from the crossing points of $\xi(T)/L$
data for different $L$ (with $\xi\equiv\xiQ$ or $\xi\equiv\xiSG$),
according to the scaling law $\xi/L=f(L^{1/\nu}(T-\Tg))$, where $f(x)$
is a scaling function and $\nu$ is the correlation length exponent
\cite{mc_ea}.


\begin{figure}[!t]
\includegraphics[width=0.47\textwidth,clip=true]{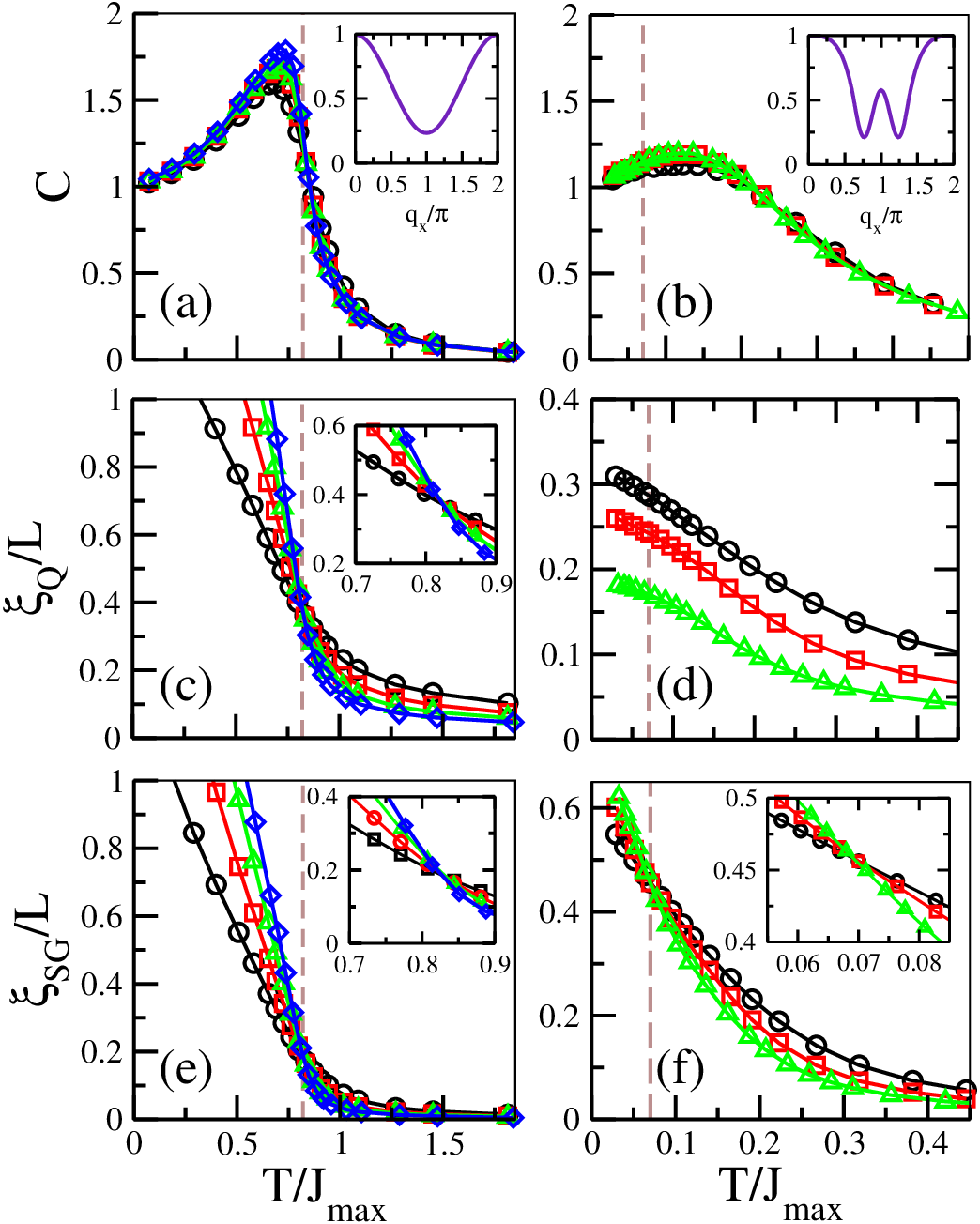}
\caption{
Classical MC results for the
ordering process of impurity moments described by $\mathcal{H}_{{\rm eff}}$ \eqref{heff},
in the cases of commensurate (left) and incommensurate (right) bulk correlations, for
$x=2\%$ and different system sizes: $L=16\left(\Circle\right)$,
$24\left({\color{red}\square}\right)$, $32\left({\color{green}\triangle}\right)$, and
$40\left({\color{blue}\lozenge}\right)$.
(a, b) Specific heat $C$ as a function of $T$.
Insets: Host dispersion $\varepsilon\left(\vec{q}\right)$,
normalized by its bandwidth, along $\vec{q}=\left(q_{x},\pi,\pi\right)$
and $\vec{q}=\left(q_{x},\frac{3}{4}\pi,\frac{3}{4}\pi\right)$, respectively.
(c, d) Magnetic correlation length $\xiQ$ divided by $L$ as function
of $T$ for $\vec{Q}=\left(\pi,\pi,\pi\right)$ and
$\left(\frac{3}{4}\pi,\frac{3}{4}\pi,\frac{3}{4}\pi\right)$,
respectively.
(e, f) $T$ dependence of the spin-glass correlation
length $\xiSG/L$. The crossing point for different $L$ defines
the ordering/freezing temperature $\Tg$ (vertical dashed lines).
Insets in c, e, f show a zoom on $\xi/L$ near $\Tg$.
Error bars are smaller than the symbol size.
}
\label{fig:comp}
\end{figure}

\ifepl
\section{Spin-glass order with short-range correlations}
\else
\textit{Spin-glass order with short-range correlations.}
\fi
A detailed comparison between the cases of commensurate and IC host correlations is in
Fig. \ref{fig:comp}. We have used the bulk triplon dispersions
$\varepsilon\left(\vec{q}\right)$ in Figs. \ref{fig:comp}a and b, respectively, to
generate the $J_{ij}$ according to Eq.~\eqref{jij}. Fig.~\ref{fig:comp}a has a single
minimum at $\left(\pi,\pi,\pi\right)$ whereas Fig.~\ref{fig:comp}b has eight minima at
$\left(\pi\pm\delta,\pi\pm\delta,\pi\pm\delta\right)$ with $\delta=\pi/4$ \cite{ic_note}.
In both cases we have used $x=2\%$ and chosen $\varepsilon\left(\vec{q}\right)$ such that
$\xib\simeq3a$.
Energies are given in units of $\Jmax=|J_{ij}(r_{ij}=1)|$ \cite{Jmax}.

All features in Figs. \ref{fig:comp}a,c,e
consistently point towards the onset of long-range magnetic order
at $\TN\approx0.82\Jmax$: a unique crossing point in both $\xiQ/L$
and $\xiSG/L$ at the same $\TN$ \cite{harris} and a specific-heat
maximum which both sharpens and moves towards $\TN$ upon increasing
$L$.

\begin{figure}[!t]
\includegraphics[width=0.49\textwidth,clip=true]{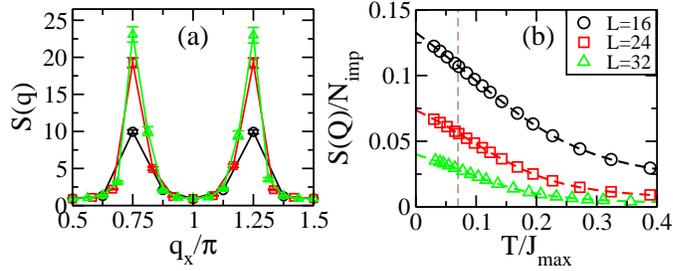}
\caption{
Structure factor $S(\vec{q})$ for IC bulk correlations and the same parameters as in Fig.
\ref{fig:comp}.
a) Momentum dependence along $\vec{q}=\left(q_{x},\frac{3}{4}\pi,\frac{3}{4}\pi\right)$
at $T=T_{g}/2$.
b) $S(\vec{Q})$ at
$\vec{Q}=\left(\frac{3}{4}\pi,\frac{3}{4}\pi,\frac{3}{4}\pi\right)$
divided by $\Nimp$ as function of $T$.
}
\label{fig:Sq}
\end{figure}

The results for the IC case, Figs.~\ref{fig:comp}b,d,f, are strikingly
different. A crossing is present only in $\xiSG/L$, indicating a
transition into a spin-glass state at a low $\Tg\approx0.07\Jmax$ \cite{critsg},
while no crossing is observed in $\xiQ/L$. The peak in $C(T)$ is
broad and occurs at a temperature considerably larger than the freezing
temperature (here $T_{peak}\approx2\Tg$). The origin is the emergence
of short-range magnetic order far above $\Tg$ \cite{fischer},
which is nicely visible in the gradual increase of $\xiQ$, Fig.~\ref{fig:comp}d.
The momentum dependence of the structure factor $S(\vec{q})$, Fig.~\ref{fig:Sq}a,
indeed shows peaks at the wavevectors $\vec{Q}$ corresponding to
the minima of $\varepsilon\left(\vec{q}\right)$. However, those peaks
grow slower than the system size, Fig.~\ref{fig:Sq}b, indicating
static short-range order with a vanishing magnetic order parameter
$M$ \cite{no_order}.

The fact that $S(\vec{q})$ peaks at (or near) $\vec{Q}$ in the IC case is non-trivial
given that the system is both strongly disordered and frustrated. It may be rationalized
by considering the dense limit $x\to1$: Here, the classical $\mathcal{H}_{{\rm eff}}$
\eqref{heff} is minimized by a spiral state with ordering wavevector $\vec{Q}$ for which
the Fourier transform of $J_{ij}$ has a maximum \cite{yosida}. Our results show
that, upon dilution, this spiral order then becomes both short-ranged and glassy.

\ifepl
\section{Doping dependence}
\else
\textit{Doping dependence.}
\fi
The model \eqref{heff} has two characteristic length scales as input, namely $\ell$ and
$\xib$, and one expects that their ratio $\ell/\xib$ is dominant in determining the
physical behavior.
Clearly, $\ell\gg\xib$ represents a strongly disordered regime, with a broad
distribution of coupling constants and suppressed ordering temperature, whereas
$\ell\ll\xib$ can be expected to lead to a more conventional state (still being
glassy in the IC case). Many experiments are in an intermediate regime of
$\ell\gtrsim\xib$ \cite{laflo}.

\begin{figure}[!b]
\includegraphics[width=0.47\textwidth,clip=true]{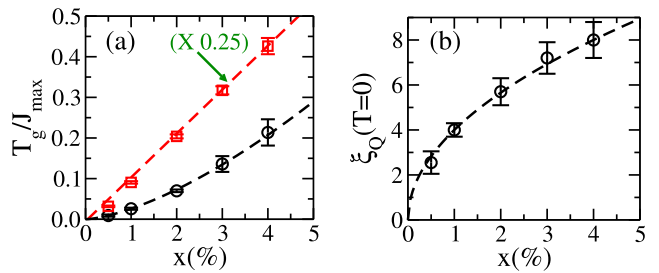}
\caption{
a) Ordering temperature (rescaled by $0.25$, ${\color{red}\square}$) for commensurate
and freezing temperature ($\Circle$) for
IC host correlations, as function of the impurity concentration $x$.
b) Magnetic correlation length $\xiQ$ for IC host correlations,
extrapolated to $T=0$ and $L=\infty$, as function of $x$.
(Dashed lines are guides to the eye, see text.)
}
\label{fig:Tg}
\end{figure}

We have calculated the ordering/freezing temperature $\Tg$ as function of doping $x$ at
fixed $\xib$, with results shown in Fig.~\ref{fig:Tg}a. As expected, $\Tg/\Jmax$ is
significantly smaller in the IC case as compared to the commensurate one due to
frustration. $\Tg(x)$ is approximately linear in this range of $x$ for the commensurate
case, while it shows sublinear behavior for the IC case, which is well fitted by $\Tg\propto x^{1.6}$ (dashed).
For smaller $\xib$ also the commensurate case shows sublinear
behavior (not shown). Experimentally, a linear $\Tg(x)$
is frequently observed \cite{laflo}, while for very small $x$ sublinear behavior has been
found \cite{manabe98}, consistent with our numerical results and previous ones for the
commensurate case \cite{laflo04,laflo}.

Thus, $\Tg(x)$ appears to be generically sublinear at small $x$ before it crosses over to
approximately linear behavior. It has been suggested that the typical value of the
couplings $J_{ij}$ governs $\Tg$ in the dilute limit $x\to0$, while the average coupling
is relevant in the more dense case where $\Tg(x)$ is found to be linear \cite{laflo,graphene_foot}.
We note, however, that at present there is no analytical understanding of the behavior of
$\Tg(x)$ in the dilute limit.

Fig. \ref{fig:Tg}b shows the magnetic correlation length $\xiQ$ in
the IC case, extrapolated to $T=0$ and $L=\infty$. Remarkably, $\xiQ(x)$
appears to vanish as $x\to0$, i.e., the spatial correlations between
the impurity moments diminish in the dilute limit.
Qualitatively, such a $\xiQ\ll\ell$ corresponds weakly correlated isolated impurities,
whereas $\xiQ\gtrsim\ell$ can be associated with a spin glass formed
from clusters of correlated impurities with cluster size $\xiQ$.
(For the parameters in Fig. \ref{fig:Tg}b, $\xiQ\approx\ell$ at $x\approx1\%$.)


\ifepl
\section{Dimensionality}
\else
\textit{Dimensionality.}
\fi
We have also considered spatially anisotropic
cases where the IC host correlations only extend along one or two
directions (dubbed 1dIC and 2dIC), whereas the correlations are commensurate
(and weaker) in the remaining direction(s) -- this situation
applies to various quasi-1d or quasi-2d materials.

\begin{figure}[!t]
\includegraphics[width=0.48\textwidth,clip=true]{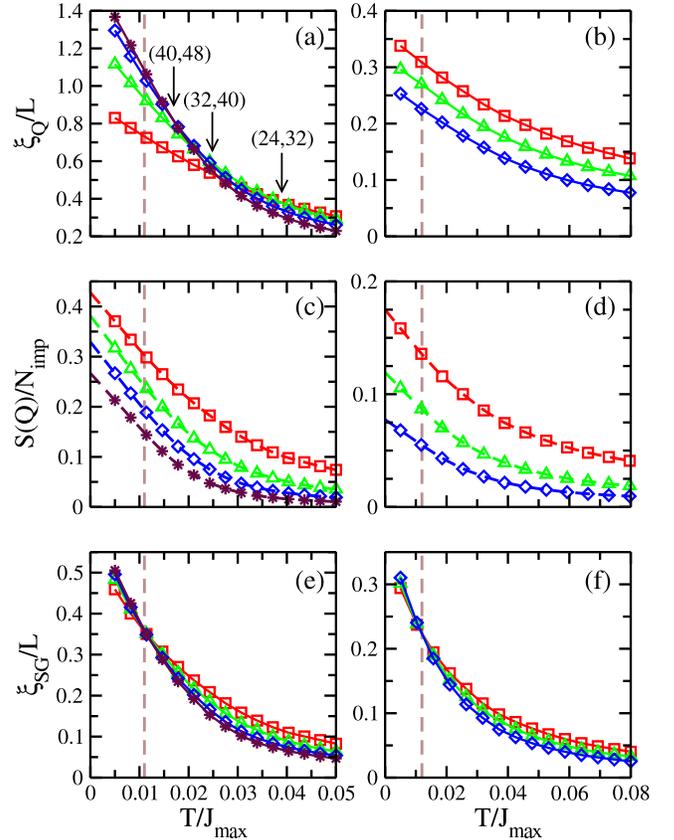}
\caption{
Ordering process of impurity moments, now for spatially anisotropic incommensurability,
comparing 1dIC with $\vec{Q}=\left(\frac{3}{4}\pi,\pi,\pi\right)$ (left) and 2dIC with
$\vec{Q}=\left(\frac{3}{4}\pi,\frac{3}{4}\pi,\pi\right)$ (right).
The bulk correlation lengths are $\xib/a\approx(4,1.5,1.5)$ (left) and
$\xib/a\approx(3.3,3.3,1.3)$ (right) in the three directions.
(a,b) Magnetic correlation length $\xiQ(T)/L$ along an incommensurate direction for
$x=0.5\%$ and system sizes $L=24\left({\color{red}\square}\right)$,
$32\left({\color{green}\triangle}\right)$, $40\left({\color{blue}\lozenge}\right)$, and
$48\left({\color{maroon}\ast}\right)$. In (a), the arrows indicate the crossing points of the curves for
different pairs of $L$.
(c,d) Structure factor $S(\vec{Q})/\Nimp$.
(e,f) Spin-glass correlation length $\xiSG(T)/L$.
For both 1dIC and 2dIC, a clear crossing point in $\xiSG/L$ is observed which allows to
extract the $\Tg$ (vertical dashed). For a discussion see text.
}
\label{fig:1dIC}
\end{figure}

The sample results in Fig.~\ref{fig:1dIC} demonstrate a clear trend with dimensionality
of the IC correlations: The 1dIC case in Fig.~\ref{fig:1dIC} (left) displays frustration
in only one direction, therefore magnetic correlations are longer-ranged and consequently
glassiness is weak. As can be seen from Fig.~\ref{fig:1dIC}a, $\xiQ$ exceeds our largest
system size at low $T$. However, from both the shift of the crossing points with system
size (Fig.~\ref{fig:1dIC}a), i.e., the absence of a unique crossing point, and the fact
that $S(\vec{Q})/\Nimp$ extrapolates to zero as $L\to\infty$ we can safely conclude
that there is no conventional magnetic order, but only spin-glass order at low $T$.
Turning to the 2dIC case in Fig.~\ref{fig:1dIC} (right), we observe that this
qualitatively resembles the situation in 3d, where $\xiQ$ remains small below $\Tg$.

We have studied the systematics of $\Tg(x)$ in the spatially anisotropic cases, and find
the general trend -- sublinear behavior at small $x$ and linear behavior at larger $x$ --
to be obeyed here as well. Interestingly, $\Tg(x)/\Jmax$ is only suppressed by factors
1.3---4 by 1dIC correlations as compared to the fully commensurate case (for correlation
lengths as in Fig.~\ref{fig:1dIC}a and $0.5\%\leq x \leq 2\%$); this can be contrasted to the
suppression by a factor of 10 or more found for the 3d incommensurate situation,
Fig.~\ref{fig:Tg}a.


\ifepl
\section{Quantum effects and real-space renormalization group}
\else
\textit{Quantum effects and real-space renormalization group.}
\fi
Our MC simulations presented so far neglect quantum fluctuations of the impurity moments.
(Quantum effects in the {\em bulk} are crucial for the moment formation; those are
taken into account in writing down Eq.~\eqref{heff}.)
A central question is whether the classical order of the $\vec{S}_{i}$ survives for
$S=1/2$. For the commensurate case, both real-space renormalization group (RSRG) studies
of Eq.~\eqref{heff} \cite{rsrg1,rsrg2} and numerical simulations of vacancy-doped quantum
paramagnets \cite{laflo,wessel01,yasuda01} show that the answer is yes. In particular,
the RSRG shows the generation of large spins in the course of the RG, which
indicates magnetic order at low $T$ (instead of a random-singlet ground state) -- the
same has been found for random Heisenberg models \cite{rieger}.

We have implemented the RSRG for the model in Eq.~\eqref{heff} with quantum spins 1/2 following
Refs.~\onlinecite{rsrg1,rsrg2,rieger}. In this procedure, we consider all possible pairs
of spins in the system and calculate the energy gap between their ground state and the
first excited multiplet. We then select the pair $(\vec{S}_1,\vec{S}_2)$ with the largest
energy gap $\Delta_{12}$ (which may then be identified with the system temperature). One
RG step is characterized by the decimation of the pair $(\vec{S}_1,\vec{S}_2)$ according
to the rules in Refs.~\onlinecite{rsrg1,rsrg2,rieger}, where the pair is either replaced
by a single effective spin (representing a spin cluster) or removed completely (if the
ground state of the pair was a singlet). We iterate the decimation procedure up to the
last effective spin, or until we decimate the last spin singlet.

\begin{figure}[!t]
\includegraphics[width=0.49\textwidth,clip=true]{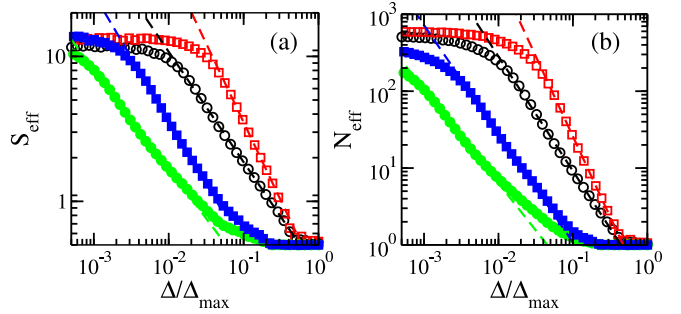}
\caption{
RSRG results for the model \eqref{heff} with quantum spins $1/2$, both for $x=1\%$, $L=50$
($\Circle$,${\color{green}\CIRCLE}$) and $x=2\%$, $L=40$ (${\color{red}\square}$,${\color{blue}\blacksquare}$),
with $\Nr=10^4$. Open (full) symbols correspond
to commensurate (IC) bulk correlations, with parameters as in Fig.~\ref{fig:comp}.
(a) Size of the effective cluster spin moment $S_{\rm eff}$, averaged over the remaining
(active) cluster spins, as a function of the energy scale $\Delta$ (equivalent to temperature),
normalized by its maximal value.
(b) Number of original spins-1/2 per cluster, averaged over the remaining
(active) cluster spins, as a function of $\Delta/\Delta_{\rm max}$.
The dashed lines are power-law fits, and the plateaus at small $\Delta$ arise
from the finite value of $N_{\rm imp}$.
}
\label{fig:rsrg}
\end{figure}

Selected RSRG results for the model \eqref{heff}, Fig.~\ref{fig:rsrg}, demonstrate the
formation of large spins upon decimation for {\em both} commensurate and IC bulk
correlations, very similar to the results in Refs.~\onlinecite{rsrg1,rsrg2,rieger}. Both
the average spin size $S_{\rm eff}$ and the average number of original spins $N_{\rm
eff}$ of the remaining spin clusters diverge in the course of the RG, indicating magnetic
order. This divergence can be described as $S_{\rm eff} \propto N_{\rm eff}^{\zeta}$
\cite{rieger}, and we estimate $\zeta$ as 0.51 for commensurate and 0.58 for IC bulk
correlations, respectively.
Importantly, the formation of a random-singlet state would instead imply that both $S_{\rm
eff}$ and $N_{\rm eff}$ remain small.
The main difference between the commensurate and IC cases in Fig.~\ref{fig:rsrg} is that
the high-temperature region where such random-singlet formation dominates is much
extended for IC correlations, reflecting a broad regime of slow fluctuations above $\Tg$,
as anticipated above. Finally, we note that the RSRG procedure is not able to distinguish
between conventional long-range magnetic order and spin-glass order, because the effect
of frustration -- which requires at least three spins -- cannot be fully captured within
the pairwise decimation procedure.

We conclude that the RSRG confirms the tendency of Eq.~\eqref{heff} toward low-temperature
order: Quantum effects do not qualitatively change the predictions of the
classical MC simulations, although they are important in a regime above $\Tg$ where the
formation of random singlets dominates in the dilute limit $\ell\gg\xib$. Hence, the
classical MC simulations cannot be expected to describe the thermodynamics at and above
$\Tg$, but they capture the physics of the low-$T$ ordered state, with our $\Tg$ being an
upper bound to the true freezing temperature.


\ifepl
\section{Experiments}
\else
\textit{Experiments.}
\fi
\bcpo\ is a spin-ladder material with 1dIC correlations \cite{tsirlin10}. Hence, we
expect weakly glassy behavior upon Zn doping \cite{laflo,alexander10}. Indeed, deviations
between experimental data and the results of numerical simulations (which do not account
for frustration) have been observed \cite{alexander10} and tentatively assigned to the
effect of magnetic frustration.
The fact that the experimental $\Tg$ of \bcpo\
is not very different from that of doped commensurate spin-gap magnets with similar
energy scales and correlation lengths \cite{laflo} is consistent with our finding that 1dIC
correlations only cause a moderate suppression of $\Tg$. From our results we predict a
strongly sublinear $x$ dependence of $\Tg$ for $x<1\%$.

The superconductor \ybca\ displays a sizeable spin gap, with 2\% of Zn substitution
inducing static short-range magnetic order peaked at a 2dIC wavevector close to that of
the host correlations \cite{ybco}.
For cuprates, experiments indicate that both IC
magnetic correlations and IC magnetic order (the latter occurs at lower oxygen doping in
Zn-free \ybco\ as well as in other cuprates) can be described as collective magnetism (as
opposed to simple Fermi-surface nesting) originating from the tendency towards stripe
formation \cite{mvrev}.
Therefore it is reasonable to assume that the physics of impurity moments in spin-gapped
cuprates can be captured by $\mathcal{H}_{{\rm eff}}$ \eqref{heff} \cite{vbs00}, and
consequently we predict the magnetically ordered state in Zn-substituted \ybca\ to be glassy.
(In fact, Ni substitution in \lsca\ has been shown to induce spin-glass order even in the
superconducting state \cite{lsco}.) Based on Fig.~\ref{fig:Tg}b we also predict the width
of the magnetic Bragg peaks to increase for smaller $x$ which can be checked by future
neutron scattering experiments.


\ifepl
\section{Conclusions}
\else
\textit{Conclusions.}
\fi
We have investigated the magnetism of impurity moments in incommensurate spin-gap
magnets, which we have shown to be frustrated, leading to static order at low $T$ akin to
that of a cluster spin glass -- in strong contrast to the commensurate case where the
low-$T$ state is a disordered, but unfrustrated antiferromagnet.

Our results call for further experiments on impurity-doped magnets, in particular a
characterization of spin-glass behavior of the low-$T$ order, e.g., by ac
susceptibility measurements.
Theoretically, it will be interesting to study the interplay of spin-glass and quantum
effects beyond the real-space RG procedure. Particularly fascinating will be the
connection of the zero-field spin glass to the Bose-glass phase expected in an applied
field \cite{rosc1}.



We thank P. Henelius, B. Keimer, C. R\"uegg, and C. Timm for illuminating discussions.
This research was supported by the DFG through FOR 960 and GRK 1621.




\begin{thebibliography}{99}
\bibitem{lin_foot} Strictly, the linear-response assumption is not
justified in deriving $J_{ij}$, as the impurity-induced moments couple
to the bulk with $J_{0}$, i.e., a relative strength of order unity.
While this is relevant for prefactors of $J_{ij}$, it will not change
its qualitative dependence of $r_{ij}$.
\bibitem{1ddiv}
In one dimension, one can show that the distribution of the $J_i^{\rm max}$ in model
\eqref{heff} displays a power-law divergence in the dilute limit, with an $x$-dependent
exponent \cite{sigrist96}.
\bibitem{critsg}
A study of the critical behavior near the putative spin-glass transition is beyond
the scope of this work. We note that the correlated disorder present in model
\eqref{heff} might render it different from that of the E-A spin glass model.
\bibitem{xi_note}
$\xiQ$ reflects equal-time correlations of the static impurity
moments, while $\xib$ parametrizes the $\w=0$ bulk susceptibility. In particular,
$\xiQ<\xib$ is possible and corresponds to frustration (i.e. destructive interference)
between the polarization clouds around each impurity \cite{laflo,alexander10,vbs00}.
\bibitem{harris}
We expect the magnetic transition to be in the 3d Heisenberg universality class even in
the presence of disorder, as the correlation length exponent $\nu\approx0.71$ fulfills
the Harris criterion $\nu>2/d$. However, from the data in Fig.~\ref{fig:comp}c we extract
a different $\nu\approx0.79$ which we attribute to large subleading corrections for our
small systems.
\bibitem{no_order}
In the IC case, we did not detect any order in scalar or vectorial spin correlations.
\end{thebibliography}
\end{document}